\documentclass[sigconf]{acmart}

\AtBeginDocument{%
  \providecommand\BibTeX{{%
    \normalfont B\kern-0.5em{\scshape i\kern-0.25em b}\kern-0.8em\TeX}}}

\copyrightyear{2021}
\acmYear{2021}
\acmDOI{}

\acmConference{}
\acmBooktitle{}
\acmPrice{}
\acmISBN{}

\usepackage{longtable}
\usepackage{multicol}
\usepackage{ltablex}
\usepackage{caption}
\usepackage{subcaption}
\usepackage{multirow}
\usepackage{setspace}

\usepackage{titlesec}

\begin{document}

\title{On the Feasibility of Predicting Questions being Forgotten in Stack Overflow}

\author{Thi Huyen Nguyen, Tu Nguyen, Tuan-Anh Hoang, Claudia Niederée}
\affiliation{%
  \institution{L3S Research Center}}
\email{{nguyen,tunguyen,hoang,niederee}@l3s.de}





\begin{abstract}
  
For their attractiveness, comprehensiveness and dynamic coverage of relevant topics, community-based question answering sites such as Stack Overflow heavily rely on the engagement of their communities: Questions on new technologies, technology features as well as technology versions come up and have to be answered as technology evolves (and as community members gather experience with it). At the same time, other questions cease in importance over time, finally becoming irrelevant to users. Beyond filtering low quality questions, “forgetting” questions, which have become redundant, is an important step for keeping the Stack Overflow content concise and useful. In this work, we study this managed forgetting task for Stack Overflow. Our work is based on data from more than a decade (2008 - 2019) - covering 18.1M questions, that is made publicly available by the site itself. For establishing a deeper understanding, we first analyze and characterize the set of questions about to be forgotten, i.e., questions that get a considerable number of views in the current period but become unattractive in the near future. Subsequently, we examine the capability of a wide range of features in predicting such forgotten questions in different categories. We find some categories in which those questions are more predictable. We also discover that the text based features are surprisingly not helpful in this prediction task, while the meta information is  much more predictive.
\end{abstract}



\keywords{Stack Overflow, forgetting management, questions deletion}


\maketitle

\section{Introduction}
{\bf Motivation}. Stack Overflow\footnote{https://stackoverflow.com/} is one of the most popular websites for users to ask and answer questions about topics in software development. Since its advent in 2008, the site has been heavily used by both professional software developers and enthusiastic programmers to share and deepen their knowledge. As of September 2019, Stack Overflow has attracted over 10.9M users and 18.1M questions. The site volume is still growing with more than 150K new questions created every month\footnote{https://data.stackexchange.com/}. 

The rapid growth of Stack Overflow website brings challenges for managing its content and keeping it attractive. Given its huge data set (in terms of questions), manual management is very costly and exhausting for the site's moderators \cite{moder2019, workload2019}. Thus, there is an emergent need for automatic methods that support the effective management of the site's content. This need has attracted the attention of different researchers in recent years \cite{Ashton2012, anton2014, Sebastian2019}. 

Despite a number of existing works on Stack Overflow, there are only few targeting automated management and cleaning of its data. Also, most of previous works focus on detecting and filtering low quality or duplicated questions \cite{Muhammad2016, denzil2013, Denzil14}. In this research, we postulate that it makes sense to even discard high quality questions once they are not relevant to users' collective information need any more, e.g. because the question relates to an outdated version of a software or a library. Keeping these irrelevant questions, on one hand, increases the cost of storing and retrieving data, while creating no or little value. On the other hand, the presence of unnecessary information affects the user experience, which may cause the dissatisfaction of users and drive them away from the platform. 

\begin{figure}[t]
    \centering
    \includegraphics[scale=0.37]{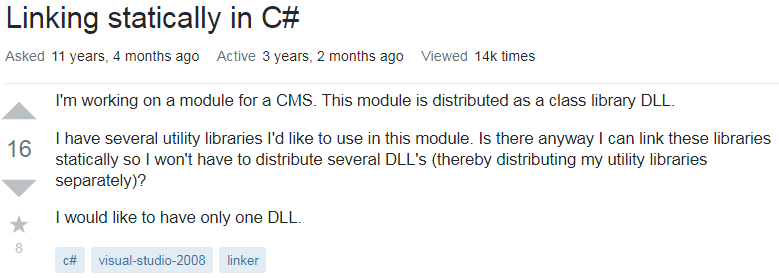}\\
    \includegraphics[scale=0.5]{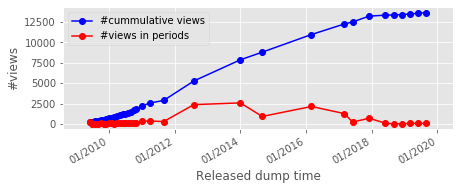}
    \caption{An example of a question being forgotten}
    \label{fig:ForgottenExample}
\end{figure}
We propose the idea of "forgetting" such unattractive questions, i.e. to let them sink away from users perception in a very similar way as irrelevant memories are forgotten. 
As a first step, in this work, we approach the problem of identifying such quality-but-irrelevant content in Stack Overflow by investigating the early identification of questions that are originally frequently encountered but will become less important and fade away over time. Precisely, those consist of questions that were often asked and (1) highly viewed recently by many users facing similar issues, but then (2) will attract only few views in the future. We call those \emph{forgotten questions}. We show an example of a forgotten question in Figure \ref{fig:ForgottenExample}. This question was created in 2008 and highly viewed until late 2017. Specifically, it got 710 views in last 6 months of 2017. However, it attracted only few views during recent 6 months. 
\begin{figure*}[!h]
    \centering
    \includegraphics[scale=0.5]{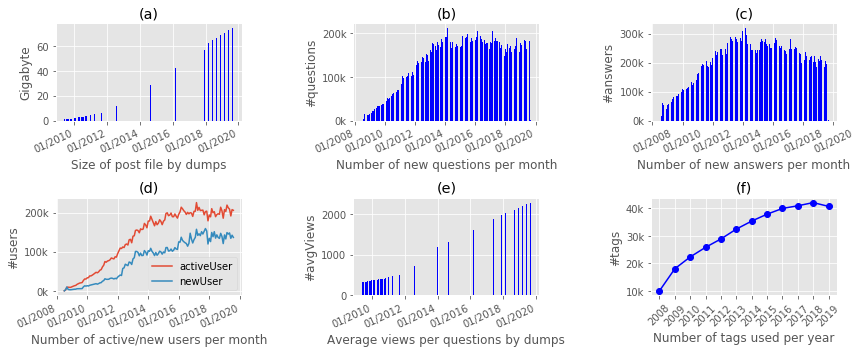}
    \caption{The dataset volume over time until September 2019\label{fig:datasize}}
    \vspace{-0.5cm}
\end{figure*}

{\bf Research Objectives}. Our analysis shows that questions are not forgotten rapidly, but in a slow pace. That is, in most cases, it takes years for a highly viewed question to become forgotten. Moreover, the prediction for far future is neither accurate nor practical. We therefore focus on questions that are in the process of being forgotten in the near future. Precisely, given a question that is highly viewed in recent period, we want to predict if the question's number of views will decrease significantly in next period. For instance, in the example shown in Figure \ref{fig:ForgottenExample}, at time point 12/2017, where the question was highly viewed, instead of predicting if the question will be forgotten after 2 years, we would like to predict if its number of views will drop down in the next 3 or 6 months.

Our work here is motivated by the idea of managed forgetting~\cite{claudia2018}: items of decreasing importance are unstressed for improving the focus on the important things - very similar to the role of forgetting in human cognitive processes~\cite{David2007}. Particularly, we would like to examine factors suggested by prior literature and general intuition that might affect the process of being forgotten of high quality questions in Stack Overflow. Furthermore, we also develop a framework for early detection of such questions. This forgetting management framework can be a good step to guide moderators to pay more attention and take further actions for unnecessary content. It also helps to bring more important information to users and encourage user engagement. For example, experienced developers would be more interested in recently highly viewed questions and are motivated to interact with them. Subsequently, this helps to improve the rank of these questions by search engines and makes it easier for users to search for information they currently need.

It is important to emphasize that our problem is different from and complements other content-driven works such as \cite{Haoxiang2019}, which examines questions' answers for detecting obsolete content within questions. Such content is either no longer correct at creation time or is becoming incorrect in near future.
In contrast, the forgetting of questions can be triggered by many reasons, in which becoming incorrect or outdated is just one of them. 

{\bf Contributions}. We summarize our main contributions in this work as follows.
\begin{itemize}
\item We present the first study of forgetting management on Stack Overflow. We show that the forgetting process is prevalent among high quality questions though at rather slow pace.
\item We investigate a wide range of factors that may have effect on questions being forgotten, and examine their capability as predictors for questions being forgotten.
\item We examine the performance of several prediction models in identifying questions that are being forgotten. We identify some categories in which the questions being forgotten are more predictable though the prediction task is generally hard. We also show the predictiveness of different feature groups and analyze their relative importance in those models.
\end{itemize}

The rest of the paper is organized as follows: Section 2 discusses related work. Section 3 presents a description of the used Stack Overflow dataset as well as our study of forgotten questions. We then conduct experiments using a machine learning framework to predict forgotten questions on Stack Overflow in Section 4. We discuss some factors that can influence our forgetting prediction task in Section 5. Finally, we summarize our main contribution and discuss some promising directions for future work in Section 6.
\section{Related Work}
\label{sec:relatedWork}
In this section, we briefly review prior works on content management in Stack Overflow that are closely related to ours. Those include works on (i) identifying low quality questions and (ii) detecting \emph{closed} or \emph{deleted} questions.

Yao et. al. \cite{Yao2013} conducted the first investigation of factors that affect the quality of questions and answers. Here, the quality is measured by scores - the difference between number of up votes and down votes the questions got \cite{score}. It is noticed that the quality of an answer is highly correlated with that of its question. Hence, the score of a question is a good predictor for predicting the quality of an answer and vice versa. Some other factors that highly influence questions/answers' quality are user-based features and community based values such as number of comments, answers, favorites of the questions within one day of questions' creation time. Later, there were several followed-up works on detecting low-quality questions using text-based or user-based features \cite{Luca2014, Piyush2015, haifa2016}. These works mostly concern with website quality maintenance by detecting low quality content at their creation date. Meanwhile, we also would like to maintain website quality but concentrate more on user information need. Nevertheless, features used in those papers can be useful for our predictions as they can well characterize different types of questions.

Correa et. al. \cite{denzil2013} presented the first study of \textit{closed} questions on Stack Overflow. These questions are either \textit{duplicate, off-topic, unclear when asking, too broad or subjective} \cite{closedQuestion}. The authors characterized different categories of \textit{closed} questions and showed that the decision to close a question mainly relies on moderator's intervention. This work also proposed a model for predicting \textit{closed} questions using four feature groups:  \textit{User Profile, Community-Process, Question Content} and \textit{Textual Style}. Following this work, Zhang et. al. \cite{Wei2017, wei2017j} proposed models to concentrate on detecting one type of \textit{closed} questions - \textit{duplicate} questions. Later on, Correa et. al. gave another study on prediction of \textit{deleted} questions \cite{Denzil14}. The authors developed feature groups in previous work to detect questions which are extremely off-topic or very poor in quality and can be deleted from Stack Overflow. Again, these papers work on detecting  low-quality content at their creation time for further quality maintenance. Intuitively, among our originally useful questions, \textit{closed} and \textit{deleted} ones are more likely to be forgotten. Then, we would want to examine the predictiveness of features in these works in our prediction task.

Our work adopt the concept of managed forgetting which is first inspired by research in psychology \cite{anderson1994, Gesine2014}. In these works, the authors investigated the forgetting behavior of human brain and its capacity to focus on important things. Managed forgetting approach in cognitive science was further broaden and related to computer science methods. Gorfein et. al. \cite{David2007} gave the concept of inhibition, which brings the idea of lowering irrelevant items and enhancing important content to forgetting in information technology. Following this idea, human-inspired digital forgetting was studied in several works \cite{claudia2015, claudia2018}. The authors suggest that information representation and assessment are important steps to determine the importance of information. It was then applied in analyzing relevant factors for content retention in social media \cite{kaweh2014} and predicting the importance of photos for retention \cite{Andrea2018}. These works give us general intuition and suggest potential factors for forgetting questions on Stack Overflow, such as dependencies from decreasing topic relevance, the existence time and quality of questions, etc., which we investigate in this work.

\section{Data Analysis}
\begin{figure}[!t]
\begin{center}
\includegraphics[scale=0.33]{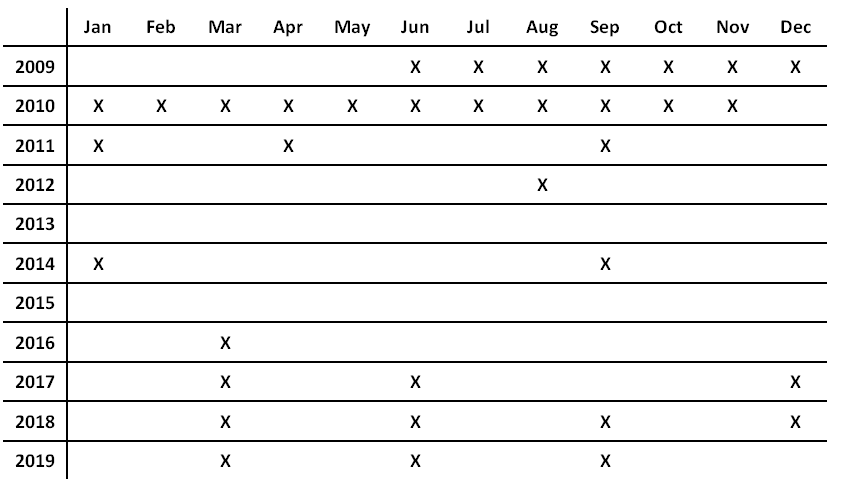}
\caption{List of data dumps provided on Stack Overflow website between 2009 and 2019}
\label{fig:statistics}
\end{center}
\end{figure}

In this section, we present our analysis in detail. We start by introducing the datasets provided by Stack Overflow. Base on analysing the datasets, we then state the formal setting for the problem of predicting questions being forgotten. Lastly, we describe our investigation on predictive features for the problem.
\subsection{Data Description}
\label{dataDescription}


Stack Overflow has been publishing its whole user-generated content since its foundation in 2008. The data is published in form of dumps under cc-wiki license \cite{license2009}, and has been regularly updated every few months \cite{dumps2014}. Figure \ref{fig:statistics} outlines the list of data dumps published up to September 2019. Each dump contains files of comprehensive information for questions, answers, comments, users, and tags. Figure \ref{fig:datasize}.a shows the size of ``Post" files which mostly contain either questions or answers. It can be seen that data dumps get much larger over time, which substantially increases the cost for managing the website content. 

We downloaded all the 35 dumps so far and obtain a dataset of 18.1M questions, 27.6M answers and 10.9M registered users. As illustrated in Figure \ref{fig:ForgottenExample} (upper), each question on Stack Overflow has a short title to summarize a specific problem in the content. When users post questions, they must include 1 to 5 tags to each question so as to describe question's topics. In addition, every question has a score which is the difference between number of up votes and down votes that the question gets. These votes indicate the questions' usefulness or dissatisfaction of users. Stack Overflow also designs a reward system that assigns to each user a reputation score to encourage users post good questions and answers \cite{userRep}. A question on Stack Overflow can get many views, which are saved and updated at every data dump.

We briefly investigate the popularity of the website by first examining the growth of its data size in terms of number of users, questions and views on the questions over time. We depict in Figure \ref{fig:datasize}.b the number of questions created in each month. The figure shows that on average, there are approximately 150K newly created questions every month. Additionally, million answers are posted monthly, as shown in Figure \ref{fig:datasize}.c.
Figure \ref{fig:datasize}.d shows the number of newly registered users and active users every month on the website. Users are considered active at a period if they have at least one activity on the website such as posting a question, answer or comment. In general, Stack Overflow is getting known by more people. In recent 4 years, the website has around 150K users signing up for new accounts and around 200K active users every month. As questions' number of views are cumulatively counted only for each data dump, we compute the average number of views per questions at different dumps, which is the total number of question views divided by the number of questions. We then observe a fast increment of average views as shown in Figure  \ref{fig:datasize}.e. Next, we examine the diversity of the site by looking at tags associated with its questions. Figure \ref{fig:datasize}.f depicts the total number of tags used in each year. It is shown that questions on Stack Overflow belong to a various set of topics, and the topic diversity keeps increasing by year. In summary, Figure \ref{fig:datasize}. a-f have shown that Stack Overflow has a dramatically developing tendency and becomes an important repository of knowledge.

\begin{figure}[!t]
    \centering
    \includegraphics[scale=0.45]{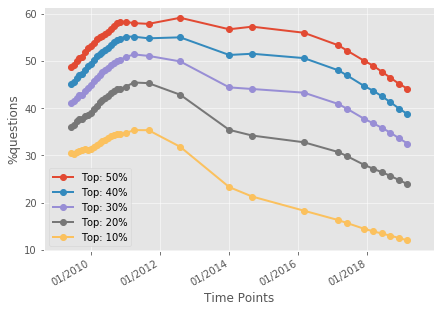}
    \caption{Percentage of questions with many accumulated views, but get less than 50 views in last 6 months[03/19-09/19]}
    \label{fig:forgottenSignal}
\end{figure}

Lastly, we examine if highly viewed questions are forgotten in Stack Overfow. To do so, for each data dump, we extract top N\% (N = 10, .., 50) questions with the highest accumulated number of views and then measure the percentage of those questions having less than 50 views in the last 6 months [03/19-09/19]. Figure \ref{fig:forgottenSignal} clearly shows that the percentage is consistently large over data dumps at different top N\%. That means, forgotten questions are prevalent among those that were highly viewed in the past. This motivates us to study the early detection of such questions in this work.

\subsection{Problem Setting}
    
     
We now develop the formal concepts of questions \textit{being forgotten} in the near future and the problem of identifying such questions among recently highly viewed ones. Generally, we rely on questions' number of views in recent period to define if they are recently highly viewed. Then we define \textit{being forgotten questions} based on the number of views in future period. Here, the periods are determined by durations between consecutive data dumps. 
\begin{table*}
\begin{center}
\caption{Datasets used in this study}
\begin{tabular}{c|c|c|c|c||c|c|c|c|c}
\multicolumn{5}{c||}{3-month-gap datasets (\#questions)}& \multicolumn{5}{|c}{6-month-gap datasets (\#questions)}\\
\hline
Current& Future&\#total&\#being&\#unforgotten&Current&Future&\#total&\#being&\#unforgotten \\

Period&Period&questions&forgotten& &Period&Period&questions&forgotten&\\
\hline
 03/18-06/18 & 06/18-09/18 &2.38M&1.17M&1.21M& 06/17-12/17 & 12/17-06/18 & 2.21M&1.50M&0.71M\\ 
 
 06/18-09/18 & 09/18-12/18&2.46M &1.08M&1.38M& 12/17-06/18& 06/18-12/18& 2.37M&1.27M&1.10M\\  
 
09/18-12/18 & 12/18-03/19&2.51M &1.38M&1.13M&
06/18-12/18 &12/18-06/19 &2.44M&1.19M&1.25M\\ 

 12/18-03/19 &03/19-06/19 &2.59M&0.99M&1.60M& 03/18-09/18 &09/18-03/19& 2.49M&1.23M&1.26M\\ 
 
 03/19-06/19&06/19-09/19 & 2.66M&1.73M&0.93M& 09/18-03/19& 03/19-09/19&2.57M&1.35M&1.22M\\
\end{tabular}
\label{tab:datasets}
\end{center}
\vspace{-0.5cm}
\end{table*}

 \begin{figure}[!t]
\begin{center}
\includegraphics[scale=0.5]{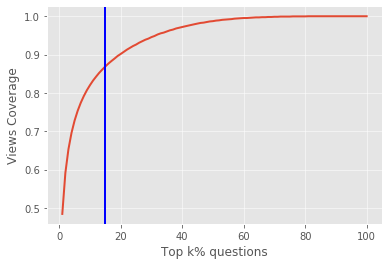}
\end{center}
\caption{Percentage of views by top K\% most viewed questions in period [03/19-09/19]. Similar patterns are observed in all other periods in our dataset.}
\label{fig:ViewIncrease}
\end{figure}

\textbf{Datasets.} We focus on questions in dumps published between June 2017 and September 2019. We then use K-month gaps to define the recent and future periods, with K = 3 or 6. That is, for a dump published at time T, the recent period of the dump is the duration of last K months up to T. Similarly, the future period is the duration from T to K months later. There are several reasons to choose those dumps and K-month settings for our forgetting study. Firstly, before June 2007, Stack Overflow did not release dumps regularly, we therefore do not have enough data to define the periods for all the dumps. Secondly, the website was not that popular in its earlier years and has grown substantially in recent years, yielding significant difference in nature between earlier dumps and recent ones. Those difference may have negative effect on the robustness of this study. Thirdly, we chose to focus on recent dumps when the website is already in its mature and stable phase in evolution so as to obtain findings that are more reusable and generalizable. Each dataset is then defined by three dumps, each is published K months after its previous dump, allowing us to compute number of questions' views in recent and future periods accurately. We only include highly viewed questions in recent period in each dataset.  

\textbf{Highly viewed questions.} To define highly viewed questions in each period, we have observed an interesting majority-minority distribution of questions' views. As illustrated by the example in Figure \ref{fig:ViewIncrease}, in every 3-month or 6-month period, top 15\% most viewed questions always get almost around than 85\% percent of total views in the periods. We therefore use the threshold of 15\% to select highly viewed questions in each dataset. When we vary this threshold around 15\%, we obtained qualitatively similar observations and results for our analysis and prediction task. Hence, for the rest of the paper, we will keep the threshold fixed at 15\%. After all, we obtain 10 large datasets for this study, as shown in Table \ref{tab:datasets}. 

\begin{figure}[t]
\begin{center}
\includegraphics[scale=0.5]{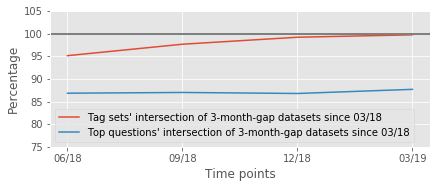}\end{center}
\caption{Percentage of highly viewed questions (respectively highly popular tags) in a  period that are still highly viewed (respectively highly popular) in the immediately next period.}
\label{fig:slowForgetting}

\end{figure}
\begin{figure}
\includegraphics[scale=0.4]{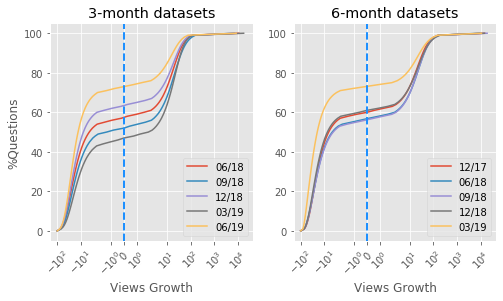}
\caption{Views growth of questions in different datasets}
\label{fig:viewsGrowth}
\end{figure}

\textbf{Question being forgotten.} Ideally, we want to predict when a highly viewed question is (mostly) forgotten. To do that, we first examine the general pace of the forgetting process by measuring the proportion of highly viewed question in a period that are still highly viewed in the immediately next period (i.e., the intersection of the two sets of highly viewed questions). Figure \ref{fig:slowForgetting} depicts this proportion across consecutive periods in our datasets. We observe that more than 85\% of highly viewed questions in current period are still be in top most highly viewed ones in the immediately next period. In addition, a large proportion of most popular tags are still highly popular in the immediately next period. Here, the popularity of a tag in a period is measured by the number of questions containing the tag posted in the period. The figure illustrates a high overlapping between popular topics of consecutive periods. All these observations show that the forgetting process of questions is rather rapid but quite slow. Hence, predicting when a question is (mostly) forgotten is a prediction for quite far future, which is neither accurate nor very practical.

We therefore do not focus on when a question is forgotten but instead on identifying questions that are in the process of being forgotten. That is, we would like to predict, for each highly viewed question in current period, if the questions' views drops significantly in the future (i.e., immediately next) period. We call such questions \textit{being forgotten questions}. Formally, to define the \textit{being forgotten questions}, we use \emph{views growth} which is defined as follows.
\begin{equation}
\centering
\label{equa:views}
    viewsGrowth(q) = \frac{futureViews(q) -  currentViews(q)}{currentViews(q)}
\end{equation}
where $currentViews(q)$ and $futureViews(q)$ is the number of views that question $q$ gets in current and future periods respectively.

Views growth values can help to observe the slow forgetting process and provide advice for moderators to take suitable actions such as deciding when to completely forget or delete questions from the website. For example, moderators can decide to delete a question if it has the dramatic drop of views in 2 continuous periods (views growth <=-90\%) or the question are forgotten in two continuous prediction periods (has no promising chance to be popular again in future). 

Figure \ref{fig:viewsGrowth} illustrates the views growth of questions in 3-month-gap and 6-month-gap datasets. Around 50\% of questions have negative views growth. That means these questions get less views in future than current period. We then define questions with number of future views less than 5\% the current views or 
$viewsGrowth < -5\%$ as \textit{ being forgotten questions}. This threshold  $-5\%$ is chosen as it is conventionally statistically significant. Also, we have slightly varied this threshold which result in qualitatively similar findings.
\begin{figure}[!t]
    \centering
    \includegraphics[scale=0.5]{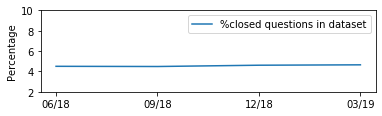}
    \caption{Percentage of \textit{closed} questions in our dataset}
    \label{fig:closePercent}
\end{figure}
\begin{figure}[ht]
    \centering
    \includegraphics[scale =0.45]{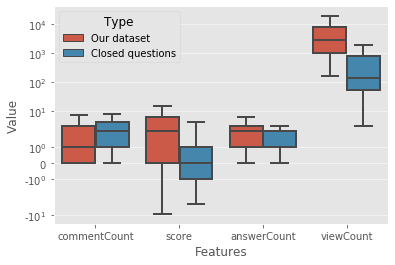}
    \caption{The difference between closed set and our dataset}
    \label{fig:closedandOurs}
\end{figure}

We make a brief observation on the quality of the questions in our datasets. Correa et. al. \cite{denzil2013} has shown that question quality on Stack Overflow has the pyramidal structure. \textit{Deleted} questions are normally extremely poor quality questions followed by \textit{closed} questions and then remaining questions. We now compare the question quality of our question set and \textit{closed} question set. Figure \ref{fig:closePercent} illustrates that only around less than 5\% of questions in our dataset are closed. Moreover, we compare other value quality indicators between \textit{closed} questions and our dataset in figure \ref{fig:closedandOurs}.  \textit{Closed} questions attract less answers, views and have smaller scores than questions in our dataset. Meanwhile, our highly viewed questions get less comments than \textit{closed} questions. This again verifies that our dataset consists of mostly high-quality questions.

\subsection{Feature Engineering}
\label{sec:dataAnalysis}


    

    

We now investigate the characteristics of \textit{being forgotten questions} by exploring a wide range of features suggested in prior works on Stack Overflow content management. Then, we examine the patterns of the features in \textit{being forgotten questions} and in those are not. Additionally, we evaluate features' capability of differentiating between the two question sets. For each feature and each dataset, assuming the dataset's periods are defined by (\emph{last dump, current dump, next dump}) triple, then the feature values are accumulated values up to the \emph{current dump}. In this section, we use the last 6-month-gap dataset (\emph{09/18, 03/19, 09/19}) for producing graphical plots as similar patterns are observed for all the other 6-month datasets. We also obtain similar patterns in 3-month-gap for most of the case though they are not always stable. This is reasonable as 3 months is considerably short span for the forgetting process to be significant. 

We categorize the features, by type of information they are engineered from, into five groups: \emph{Text-, Question-, Answer-, User-} and \emph{Tag-}based features. 

\textit{\textbf{Text-based features}}: The text in body and title and tag of questions. In non-deep learning models, we use tf-idf of text as features.

\begin{figure}[!t]
    \centering
    \includegraphics[scale=0.5]{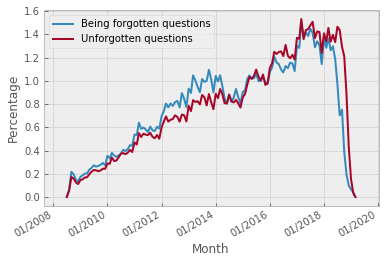}
    \caption{Percentage of \textit{being forgotten questions} and unforgotten questions over time, normalized by number of questions of each question type}
    \label{fig:forgotUnforgotOverTime}
\end{figure}

\textit{\textbf{Question-based features}} consists of the followings:
\begin{itemize}
 \item Time related features such as question age (\emph{ageByCreDate}) and question inactive time (\emph{ageByLastAct}). Question age is the number of months that the question have been existing on the website since creation time. Each question in dump has a last activity date value which is last modified time of any answer or the question itself. We compute the duration since question' last activity date till prediction time as question inactive time. Intuitively, old questions and questions that are inactive for long time are more likely to be forgotten. 
    \item Basic community-generated indicators including \emph{score, viewCount, commentCount, answerCount}. These are respectively the score, number of views, comments and answers that questions have accumulated up to the prediction time. 
    \item Questions' text based values: the numbers of characters in body (\emph{bodyLen}), code snippet (\emph{codeLen}), and title (\emph{titleLen}), and the numbers of verbs (\emph{nOfVerbs}), pronouns(\emph{nOfPRP}), and nouns(\emph{nOfNouns}) in the whole question. These features are used by most previous work \cite{denzil2013, Denzil14, Luca2014}. Features such as number of verbs, nouns, pronouns indicate if users have done certain of work before asking \cite{Vassudev2014} and then reflect the quality of questions.
\end{itemize}
\begin{figure}[!t]
    \centering
    \includegraphics[scale=0.5]{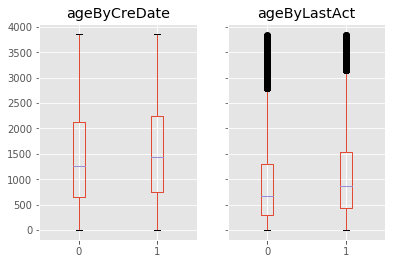}
    \caption{question's age (\textit{ageByCreDate}) and questions' inactive time(\textit{ageByLastAct}) of unforgotten questions(0), and being forgotten questions(1)}
    \label{fig:ageQuesFeature}
\end{figure}

\begin{figure}[!t]
    \centering
    \includegraphics[scale=0.5]{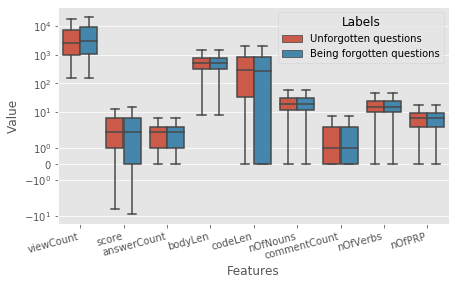}
\caption{\textit{Question-based} features of \textit{being forgotten questions} and unforgotten questions}
    \label{fig:questionsFeatures}

\end{figure}

Figure \ref{fig:forgotUnforgotOverTime} illustrates the percentage of \textit{ being forgotten questions} and unforgotten questions at different time points. \textit{Being forgotten questions} can be either newly created or old questions. In the figure, we can see that there are many highly-viewed questions created far away in the past, but continue to be popular/unforgotten in far future. In contrast, new questions with many views in current time period can be forgotten in near future. It can also be seen in Figure \ref{fig:ageQuesFeature} that \textit{being forgotten questions} tends to have higher age and unforgotten questions have shorter inactive time. In other words, unforgotten questions are  more active near prediction time. It makes sense that these questions tend to still be popular in the next period. Some other \textit{question-based} features are illustrated in Figure \ref{fig:questionsFeatures}. \textit{Being forgotten questions} get slightly more accumulated views as they exist for longer time on the website. They contain more questions with lower score and shorter code snippets. 


\textit{\textbf{Answer-based features}}: Most of questions in our data have less than 10 answers. Askers can mark an answer as accepted or the best. We expect that a question is popular in future if it gets good answers recently, hence extract features of first, best and last answers.
\begin{itemize}
    \item \emph{FirstAnsScore}, \emph{BestAnsScore}, \emph{LastAnsScore}: The scores of the first, the best and the last answer respectively. 
    \item \emph{FirstAnsComCount, BestAnsComCount, LastAnsComCount}: the numbers of comments that the first, best, and last answer gets respectively.
    \item \emph{FirstAnsBodyLen, BestAnsBodyLen, LastAnsBodyLen}: the numbers of characters in body of the first, best, and last answer respectively.
    \item \emph{TimeToGetFirstAns, TimeToGetBestAns, TimeToGetLastAns}: the numbers of minutes to get the first, best, and last answer since question's creation time.
    \item \emph{TimeToFirstAnsLastActDate}, \emph{TimeToBestAnsLastActDate}, \emph{TimeToLastAnsLastActDate}: the numbers of minutes to the first, best, and last answer activity since the answer's creation time.
\end{itemize}

\begin{figure}[!t]
    \centering
    \includegraphics[scale=0.5]{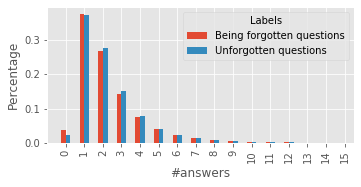}
    \caption{Number of answers of questions}
    \label{fig:answers}
\end{figure}

\begin{figure}[h]
    \centering
    \includegraphics[scale=0.5]{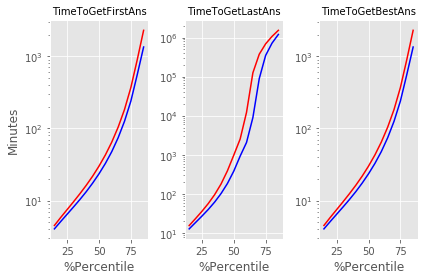}
    \includegraphics[scale=0.7]{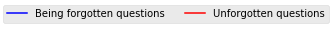}
    \caption{Time to get the first, last, and best answers}
    \label{fig:timetoanswers}
\end{figure}

Figure \ref{fig:answers} shows that unforgotten questions are more likely to be answered (e.g. having at least one answer) than ones being forgotten. Also, we have observed from Figure \ref{fig:timetoanswers} that unforgotten questions generally take longer time to get the first and best answers, and get interested for longer time than \textit{being forgotten questions}.

\begin{figure}[!t]
    \centering
    \includegraphics[scale=0.5]{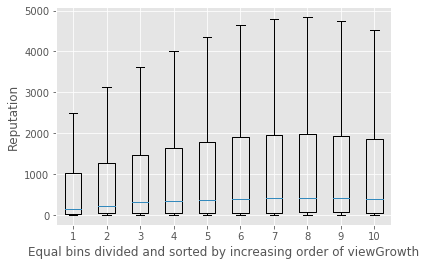}
    \caption{Reputation of authors of questions in bins of increasing viewGrowth}
    \label{fig:userRep}
\end{figure}

\textit{\textbf{User-based features}} consists of user information extracted from their profile, including:
\begin{itemize}
    \item User reputation (\emph{userRep})\footnote{https://stackoverflow.com/help/whats-reputation}: The total reward that users get when they post good questions or answers.
    \item The number of views (\emph{UserViews}): Number of times the user profile is viewed.
    \item The number of up votes (\emph{UserUpVote}), down votes (\emph{UserDownVote}) on posts that users contribute.
\end{itemize}
In existing works, the above user features are good signals to filter low quality questions from in Stack Overflow \cite{ denzil2013, Denzil14, Luca2014} as experienced users tend to create better questions so as to maintain their reputation. However, questions in our dataset are high-viewed ones, there is only a slight difference between user features of two question sets. In order to clearly see this difference between all questions in our dataset, we divide the dataset into 10 bins based on \textit{viewGrowth} and plot the distribution of user reputation in Figure \ref{fig:userRep}. The figure clearly shows that not only highly experienced users but also normal users with less reputation can create good questions that attract many views. However, questions created by lower reputation are more likely to decrease the views in future. 
\begin{figure}[ht]
    \centering
    \includegraphics[scale=0.45]{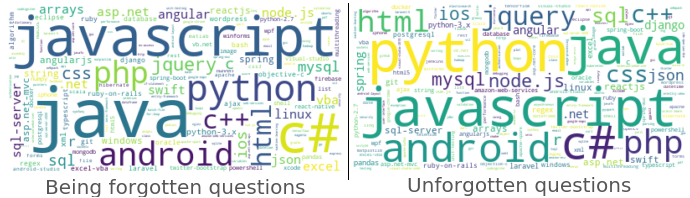}
    \caption{Top frequent tags in questions}
    \label{fig:wordcloud}
\end{figure}

\textit{\textbf{Tag-based features}}: We extract features for every tag associated with the question. Figure \ref{fig:wordcloud} depicts the word cloud of tags in our dataset. The most popular tags of both unforgotten questions and \textit{being forgotten questions} are well-known tags such as Java, Javascript, C\#, Android, etc.. That explains why these questions get many views in recent period. Nevertheless, popular topics normally evolve rapidly, people get familiar with current issues quickly and new problems are raising. Therefore, questions of popular tags are more likely to be forgotten. 

\begin{figure}[t]
    \centering
     \centering
    \includegraphics[scale=0.45]{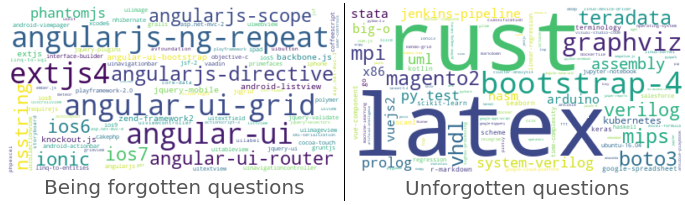}
    \caption{Top tags ranked by the ratio of number of questions containing the tag in each question set to the total number of questions of the tag in the whole data set}
    \label{fig:rankedTags}
\end{figure}

\begin{figure}[!t]
    \centering
    \includegraphics[scale=0.45]{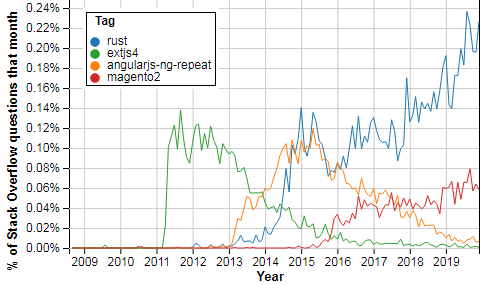}
    \caption{Example tag trends}
    \label{fig:tagTrends}
\end{figure}

We then investigate the less popular tags to examine the differences between those of unforgotten questions and \textit{being forgotten questions}. In order to have a clear observation at the trend of tags, we take questions in first 6-month dataset \textit{(06/17, 12/17, 06/18)} for visualization of this part. Firstly, we extract 1000 most frequent tags for each question set, then we rank tags in each set according to the ratio between the number of questions in the set that contains the tag and the total number of such questions in our whole dataset. The ranked tags, by this ratio, of unforgotten questions and \textit{being forgotten questions} are shown in Figure \ref{fig:rankedTags}. We observe that top ranked tags of unforgotten questions tend to have upward trends (i.e., attract more questions) in the future. In contrast, top tags in \textit{being forgotten questions} are going downward. Figure \ref{fig:tagTrends} shows the trend of two down-trending tags(\textit{extjs4, angularjs-ng-repeat}) in \textit{being forgotten question} set and two up-trending tags(\textit{rust, mangento2}) in unforgotten question set\footnote{Similar patterns of other tags can be viewed at https://insights.stackoverflow.com/trends}.

In addition, we investigate following features that show the popularity of tags. 
\begin{itemize}
    \item $TagExistTime_K$ (for $K = 1, \cdots, 5$): the number of days the tags have been existing on the website since its first use on the website.
    \item $TagPop_K$ (for $K = 1, \cdots, 5$): the number of questions using the tag up to prediction time.
    \item $TagAciveTime_K$ (for $K = 1, \cdots, 5$): the number of days the tags have used on the website since its first use to last use on the website.
    \item \textbf{$TagPrePop_{KQ}$ (for $K = 1, \cdots, 4$, $Q = 1, \cdots, 5$)}: the number of questions containing the question's $Q$th tag in $K$ previous periods before prediction time.
\end{itemize}

\textbf{Feature predictiveness}. Lastly, we examine the capability of features in separating \textit{being forgotten questions} from unforgotten ones across our 10 datasets(section \ref{dataDescription}). To do that, we conduct Man-Whitney's statistical test \cite{manwhitney}, and compute the Spearman correlation \cite{spearman} and AUC \cite{AUC} value for all features in each dataset. Generally, the test helps us to inspect the statistical relationship between two samples: whether their data points come from the same distribution. Man-Whitney test is particularly suitable for our datasets as it is  a non-parametric test. Meanwhile, Spearman correlation is also a non-parametric measure that assesses how strongly two variables are related using a monotonic function. Finally, we evaluate AUC value of every feature by running a simple model (e.g. decision tree model) using only that single feature. The results tell how much the feature is capable of distinguish between classes. 

We compute the test, correlation, AUC scores of all the features and make the following observations from the results\footnote{Please refer to Tables 2 and 3 in the extended version of this paper for the scores of all the feature: https://www.overleaf.com/read/pvhzymfjtvvg}.
\begin{itemize}
    \item All features have P-value < 0.05 for the Man-Whitney tests. That means the value of the features in being forgotten and unforgotten questions are from different distributions. 
    \item The correlations between classes (i.e., being forgotten and unforgotten) and features are weak as all the Spearman correlation are close to 0. 
    \item There are several predictive features (e.g., age of the question, time to get the first, best and last answers, the questions' active time) with AUC values at 56\%-59\%. However, all the AUC values are not greater than 60\%. That means there is no single feature that can well separate the \textit{being forgotten questions} from the unforgotten questions.  
\end{itemize}
The above results have shown that the prediction of \textit{being forgotten questions} is quite difficult. Thus, in the next section, we would like to examine how the combination of all the features can help to improve the prediction performance.
\section{Prediction}
\label{sec:prediction}



\subsection{Experiment Settings}
In this section, we examine the collective prediction performance of the feature sets by using them for training models to predict questions being forgotten. 
We tried various prediction models including Support Vector Machines, Random Forest, Gradient Boosting Trees and deep learning models. However, the deep learning models do not perform as well as we expected. This echoes the findings from recent empirical works that simpler approaches outperforms deep learning models on Stack Overflow datasets \cite{weifu2017, Bowen2018}. Overall, the Gradient Boosting Trees model \cite{gradientBoost} perform the best among all. Hence, for saving the space, we only present in the following the result of this model.

As our prediction task is a text classification problem, we first examine the performance of text based features for the task. Next, in order to measure the predictiveness of different feature groups, we add the groups incrementally and evaluate their performance on all 10 datasets. For each dataset, we perform five runs. In each run, the dataset is randomly splitted into training and test sets by ratio 90\%-10\%. In the following, we report the average performance across the runs. Here we use both accuracy (Acc) and F1 score to measure the performance. 

\subsection{Results}
Table \ref{tab:incrementalF1} shows the minimum, maximum and average performance of the sets of features on the datasets. Surprisingly, the text based features are not helpful for the task. This may be due to the fact that there are many similar questions on different versions of the same technology, but only the ones on previous versions are forgotten. Metadata features such as \textit{question-based, answer-based and tag-based} features are important features in our classification. \textit{Question-based} features return  60\% F1,  60\% Acc in the worst case and 81.21\% F1, 70.57\% Acc in the best case. The \textit{user-based} and \textit{text-based} features slightly help to improve the results of our model. This is expected as we have seen in Section \ref{sec:dataAnalysis} that all questions in our dataset are about popular topics and are created by users with similar reputation.

\begin{table*}
    \centering
    \caption{Prediction performance of different feature sets}
    \label{tab:incrementalF1}
    \begin{tabular}{l|l|l|l|l|l|l|l|l|l|l|l|l}
    \multirow{3}{*}{\textbf{Features}}& \multicolumn{6}{c|}{\textbf{3-month-gap datasets}}&\multicolumn{6}{c}{\textbf{6-month-gap datasets}}\\
    \cline{2-13}
    &\multicolumn{3}{c|}{F1 Score}&\multicolumn{3}{c|}{Acc}&\multicolumn{3}{c|}{F1 Score}&\multicolumn{3}{c}{Acc}\\
    \cline{2-13}
    &Min&Max&Avg&Min&Max&Avg&Min&Max&Avg&Min&Max&Avg\\
    \hline
     tfidf-body &1.05&78.80&41.08&55.13&65.10&58.97&41.46&80.83&60.87&53.35&67.84&57.08\\
     \hline
     tfidf-title&0.92&78.22&37.23&55.38&65.12&59.29&29.94&80.84&65.01&53.94&67.85&57.27\\
     \hline
    
    tfidf-tag&1.70&78.80&40.24&55.62&62.15&58.57&38.89&80.86&63.20&55.36&67.94&58.32\\
     \hline
     tfidf-(body+title)&1.55&78.81&40.75&55.38&65.12&59.29&46.86&80.84&65.85&54.17&67.84&57.64\\
     \hline
     Text&1.99&78.80&42.27&56.31&65.24&59.68&49.07&80.86&65.57&55.32&67.91&58.54\\
     \hline
     Question&20.34&78.79&55.40&59.88&66.84&62.40&59.55&81.21&67.83&59.81&70.57&63.58\\
     \hline
     User&0.27&78.81&37.98&52.18&65.04&58.06&25.25&80.84&54.38&51.48&67.84&56.23\\
    \hline
    Answer&8.27&78.81&47.11&53.56&65.04&59.12&50.31&80.82&62.83&54.01&67.90&57.16\\
     \hline
      Tag &3.56&78.69&47.91&56.72&65.70&60.12&52.03&80.87&63.52&55.67&68.25&59.48\\
     \hline
     Question+User&20.21&78.79&55.35&59.91&66.84&62.39&59.54&81.24&67.81&59.87&70.60&63.57\\
     \hline
     Question+User+Answer&20.10&78.82&55.65&60.02&66.91&62.58&59.66&81.30&67.95&59.98&70.69&63.78\\
     \hline
     Question+User+Answer+Tag&19.72&78.91&56.08&60.70&67.05&63.12&60.28&81.68&68.39&60.58&71.41&64.36\\
     \hline
     All&18.61&78.96&55.88&61.41&66.97&63.26&61.62&81.86&68.77&60.56&71.35&64.43\\
     \hline
    \end{tabular}
    
    \vspace{-0.5cm}
\end{table*}

\begin{figure}[!t]
        \centering
        \hspace{-0.5cm}
        \includegraphics[scale=0.5]{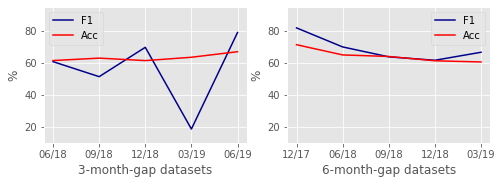}
        \caption{Prediction performance when using all features across datasets}
        \label{fig:generalResults}
    \end{figure}

Next, we would like to examine more closely the performance when using all features. Figure \ref{fig:generalResults} shows this performance across 10 datasets. Our model does not return stable performance on 3-month-gap datasets. This is reasonable as 3 months is rather a short span for the forgetting process being stable. On the other hand, we consistently obtain practically good performance for the 6-moth-gap datasets.

\begin{figure}[!t]
\begin{center}
\includegraphics[scale=0.475]{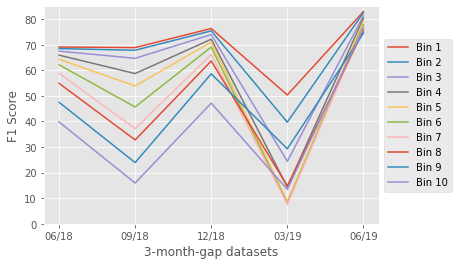}\end{center}
\caption{Results on bins of 3-month-gap datasets}
\label{fig:3monthResult}
\end{figure}

\begin{figure}[!t]
\begin{center}
\includegraphics[scale=0.475]{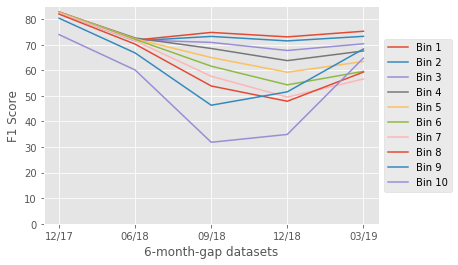}\end{center}
\caption{Results on bins of 6-month-gap datasets}
\label{fig:6monthResult}
\end{figure}

Lastly, we examine the prediction performance in different categories of questions. Since the number of recent views of questions in our dataset varies significantly, i.e. [80 - >3500] views for 3-month gap datasets or  [200 - >10000] views for 6-month gap datasets, we therefore would like to investigate how our model performs on different subsets of questions with similar views. Specifically, we divide each dataset into 10 equal bins based on number of recent views, and measure the F1-score for each bin separately. Figures \ref{fig:3monthResult} and \ref{fig:6monthResult} show these scores for 3-month and 6-month gap datasets respectively. We can observe from the figures that, across the datasets, the first bins tend to contain more \textit{questions being forgotten} and result in higher F1-scores. In contrast, questions with higher views, e.g., bins 9 or 10, tend to receive the same or higher number of views in future period. That make the datasets in these bins highly unbalanced and more difficult to predict \textit{questions being forgotten}.

\subsection{Feature Ranking}
\begin{figure}[!t]
\centering
   \includegraphics[scale=0.5]{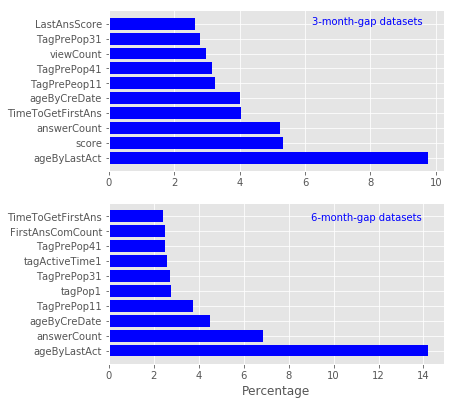}

\caption{Top features across different datasets (please refer to Section~\ref{sec:dataAnalysis} for the definition of the features)}
\label{fig:featureRanking}
\end{figure}

We now analyze the relative importance of single features in predicting the \textit{being forgotten questions}. Here the features' relative importance is computed based on their important score returned by the Gradient Boosting Trees model. For aggregating across different datasets, we take the mean of scores returned by the models trained on each dataset. We then normalize the means by their sum, and hence the relative scores of features are measured in percentage. The top 10 features with highest scores are shown for 3-month and 6-month gap dataset in Figure \ref{fig:featureRanking}. Most important features belong to \textit{question-based, tag-based} and \textit{answer-based} feature groups. This is consistent with the prediction performance presented above.  

\section{Discussion}
Despite the rich set of results we have presented above, the forgetting management of Stack Overflow questions is still a nontrivial problem. A possible reason for this hardness is that the drop of views in future period can be significantly affected by several external factors beyond our information. Examples of those factors include the competition for audience from other websites and the effect of search engines. There are many programming-oriented Q\&A websites \cite{qawebs,programmingWebs} recently launched that allow people to easily search for answers when they get stuck with their programming problem. This rich choice of resources most likely reduces the number of visits to Stack Overflow, and unpredictably change the popularity of its questions. Furthermore, the popularity of a question on webpage can be strongly affected by search engines. Very often, people look for previously asked and answered questions in Stack Overflow that are similar to theirs using some external search engine such as Google or Bing, whose working mechanism is unknown and independent to that of Stack Overflow. That makes our forgetting prediction problem become more difficult. 

Additionally, the difficulty of a question may also affect its views. For instance, an easy question normally lose their attractiveness quicker. When a topic is more popular and many people become competent at different aspects of the topic, easy questions can be sought by less people. That results in decreasing trend of the question's views. However, examining the difficulty of a question or topic changing by time is not trivial and beyond this work. Also, in seven years (2011-2017) Stack Overflow does not release data dumps frequently, we are therefore lack of information to study the trend of views and cannot extract comprehensive time series based features of questions, answers, and users. These features would help to improve the prediction performance significantly.  
\section{Conclusion}
This paper presents the first study on forgetting management of high-quality questions on Stack Overflow. We find that that many questions on Stack Overflow were highly popular in the past but drops their popularity overtime. These questions should be detected as forgotten questions for further processing such as question closure, deletion or questions' priority calculation in search or recommendation system. In general, there is no clearly different between forgotten questions and the other high-quality questions. However, forgotten questions tend to be a bit easier and take less time to get first answers and best answers. Therefore, they lose the attractiveness more quickly. We conduct a predictive model to detect forgotten questions on Stack Overflow at different time points and time gaps of 3 months or 6 months. Four main feature groups are analyzed such as questions' community values, answers' information, questions' content, tags.  The feature analysis and experiments show that each feature separately is not a good predictor. However, the combination of features can predict forgotten questions with approximately 69\% F1-score for all datasets. Furthermore, we have identify categories where the prediction performance is much higher, up to 82\% F1 score.

For the future work, we would like to examine more fine-grained factors that might affect the forgetting process of questions in Stack Overflow. These factors include the topic, difficulty, and language of the questions, as well as its relatedness to trends in technology.

\bibliographystyle{ACM-Reference-Format}
\bibliography{ref.bib}
\end{document}